 \documentclass[final,5p,times,twocolumn]{elsarticle}

\usepackage{amssymb,amsmath,hyperref}

\journal{Physica E}

\begin{document}

\begin{frontmatter}



\title{Two-dimensional InSe/WS$_2$ heterostructure with enhanced optoelectronic performance in the visible region}


\author[BTBU]{Lu-Lu Yang}
\author[PKU]{Jun-jie Shi}
\author[IMNU]{Min Zhang}
\author[ISCAS]{Zhongming Wei}
\author[PKU]{Yi-min Ding}
\author[PKU]{Meng Wu}
\author[IMNU]{Yong He}
\author[PKU]{Yu-lang Cen}
\author[PKU]{Wen-hui Guo}
\author[PKU]{Shu-hang Pan}
\author[BTBU]{Xing-Lai Che}
\author[BTBU]{Xiong Li}
\author[BTBU]{Yao-Hui Zhu\corref{cor1}}
\address[BTBU]{Physics Department, Beijing Technology and Business University, Beijing 100048, China}
\address[PKU]{State Key Laboratory for Artificial Microstructures and Mesoscopic Physics, School of Physics, Peking University, Beijing 100871, China}
\address[IMNU]{College of Physics and Electronic Information, Inner Mongolia Normal University, Hohhot 010022, China}
\address[ISCAS]{State Key Laboratory of Superlattices and Microstructures, Institute of Semiconductors, Chinese Academy of Sciences \& College of Materials Science and Opto-Electronic Technology, University of Chinese Academy of Sciences, Beijing 100083, China}
\cortext[cor1]{Corresponding author.}
\ead{yaohuizhu@gmail.com}

\begin{abstract}
Two-dimensional (2D) InSe and WS$_2$ exhibit promising characteristics for optoelectronic and photoelectrochemical applications, e.g. photodetection and photocatalytic water splitting. However, both of them have poor absorption of visible light due to wide band gaps. 2D InSe has high electron mobility but low hole mobility, while 2D WS$_2$ is on the opposite. Here, we design a 2D heterostructure composed of their monolayers and study its optoelectronic properties by first-principles calculations. Our results show that the heterostructure has a direct band gap of 2.19 eV, which is much smaller than those of the monolayers mainly due to a type-II band alignment: the valence band maximum and the conduction band minimum of monolayer InSe are lower than those of monolayer WS$_2$, respectively. The visible-light absorption is enhanced considerably, e.g. about fivefold (threefold) increase at the wavelength of 490 nm in comparison to monolayer InSe (WS$_2$). The type-II band alignment also facilitates the spatial separation of photogenerated electron-hole pairs, i.e., electrons (holes) reside preferably in the InSe (WS$_2$) layer. The two layers complement each other in carrier mobilities of the heterostructure: the photogenerated electrons and holes inherit the large mobilities from the InSe and WS$_2$ monolayers, respectively.
\end{abstract}

\begin{keyword}
2D InSe/WS$_2$ heterostructure \sep First-principles calculations \sep Electronic structures \sep Absorption spectrum \sep Mobility


\end{keyword}

\end{frontmatter}


\section{Introduction}
\label{Intro}

Photovoltaics and photocatalysis are two promising ways to convert solar energy into easy-to-use forms, i.e., electricity and fuels (e.g.~H$_2$)~\cite{Fujishima1972,Graetzel2001}. Both of the approaches need semiconductors satisfying several basic requirements, e.g. appropriate band gaps, low recombination rates of photogenerated electron-hole pairs, and high carrier mobilities. In the past decade, many researchers have turned their attention to two-dimensional (2D) semiconductors, such as group-III monochalcogenides (MX, M=Ga and In, X=S, Se, Te)~\cite{Huang2016,Xu2016}, transition-metal dichalcogenides (TMDCs) of MX$_2$ type (M=Mo, W; X=S, Se, Te)~\cite{Wang2012,Chhowalla2013}, and carbon nitrides (e.g.~g-C$_3$N$_4$ and C$_2$N)~\cite{Wang2009,Mahmood2015}. Studies on 2D semiconductors have demonstrated their advantages over the corresponding bulk materials in various ways~\cite{Xu2013}.

Recently, 2D InSe has exhibited great potential for optoelectronic and photoelectrochemical applications due to its high electron mobility, good metal contacts, and wide band-gap range~\cite{Feng2014,Feng2015,Bandurin2017}. Monolayer and/or few-layer InSe have been prepared by mechanical exfoliation~\cite{Bandurin2017,Mudd2013,Lei2014} and chemical vapor transport~\cite{Ho2016} methods from bulk InSe. The band gap increases from $1.26$ eV to $2.9$~eV as bulk InSe is thinned to a monolayer~\cite{Mudd2013,Zhuang2013}. Correspondingly, its photoluminescence has a broad response from the near infrared to the visible region~\cite{Royo2014,Mudd2015}. The electron mobility of few-layer InSe exceeds $10^3$ cm$^2$V$^{-1}$s$^{-1}$ and $10^4$ cm$^2$V$^{-1}$s$^{-1}$ at room and liquid-helium temperatures, respectively~\cite{Bandurin2017}. It is also predicted that 2D InSe is a potential photocatalyst for hydrogen production via water splitting with solar energy~\cite{Zhuang2013,Peng2017}. However, the optoelectronic properties of monolayer InSe need to be improved in several ways before it can be exploited widely in practice. First, its hole mobility is low although it has a large electron mobility. Second, its monolayer has a poor absorption of visible light due to the indirect wide band gap~\cite{Hu2017}. Finally, the recombination rate of photogenerated electron-hole pairs needs to be reduced further. One of the methods to overcome these difficulties is the construction of van der Waals (vdW) heterostructures with type-II band alignment~\cite{Ding2017,Yan2017,Ding2018,Ding2018SSC}.

The present work is aiming to enhance the optoelectronic performance of 2D InSe by combining it with another suitable 2D material to build a vdW heterostructure. One of its possible counterparts is 2D WS$_2$, which has a hexagonal structure like 2D InSe~\cite{He2016,Ju2018,Kumar2018,Wang2018}. 2D WS$_2$ belongs to the broad family of 2D TMDCs and has also brought about widespread attention among experimental and theoretical researchers recently. It has been successfully synthesized by various techniques, e.g. liquid exfoliation~\cite{Coleman2011}. Researchers have investigated its various applications, such as photocatalysis and optoelectronics~\cite{Perea2013,Zhuang2013WS2,Yang2015,Pesci2017}. Bulk WS$_2$ has an indirect band gap, and it becomes a direct band gap semiconductor with its valence band maximum (VBM) and condition band minimum (CBM) located at K points when thinned to a monolayer~\cite{Yun2012}. Furthermore, its field-effect mobility can reach 140 cm$^2$V$^{-1}$s$^{-1}$ at low temperature~\cite{Ovchinnikov2014}. We study the optoelectronic properties of the 2D InSe/WS$_2$ vdW heterostructure by first-principles calculations in this paper. The results demonstrate an enhanced optoelectronic and photocatalytic performance of the heterostructure in the visible region.

The present paper is organized as follows. Computational details are given in Sec.~\ref{sec2}. Then the results are presented and discussed in Sec.~\ref{sec3}. Finally, the main conclusions are summarized in Sec.~\ref{sec4}.

\section{Computational methods and models}
\label{sec2}

Our first-principles calculations were performed with the Vienna Ab initio Simulation Package (VASP)~\cite{Kresse1993,Kresse1996}. Within the framework of the density functional theory (DFT), we used the generalized gradient approximation (GGA) with the Perdew-Burke-Ernzerhof (PBE) functional to describe the exchange and correlation interactions between the valence electrons~\cite{Perdew1996}. The electron-ion interactions were described by the projector augmented-wave (PAW) method~\cite{Kresse1999}. In optimizing geometric structures, the vdW interaction was accounted for by the optB88 vdW exchange functional~\cite{Klimes2010,Klimes2011}. The thickness of vacuum layer is more than 20 {\AA} so as to ensure decoupling between periodically repeated layers, and the energy cutoff is set to 450~eV. The structures were fully optimized until the residual atomic forces were smaller than 0.01 eV/{\AA}.

\begin{figure}
\centering
  \includegraphics[width=0.48\textwidth]{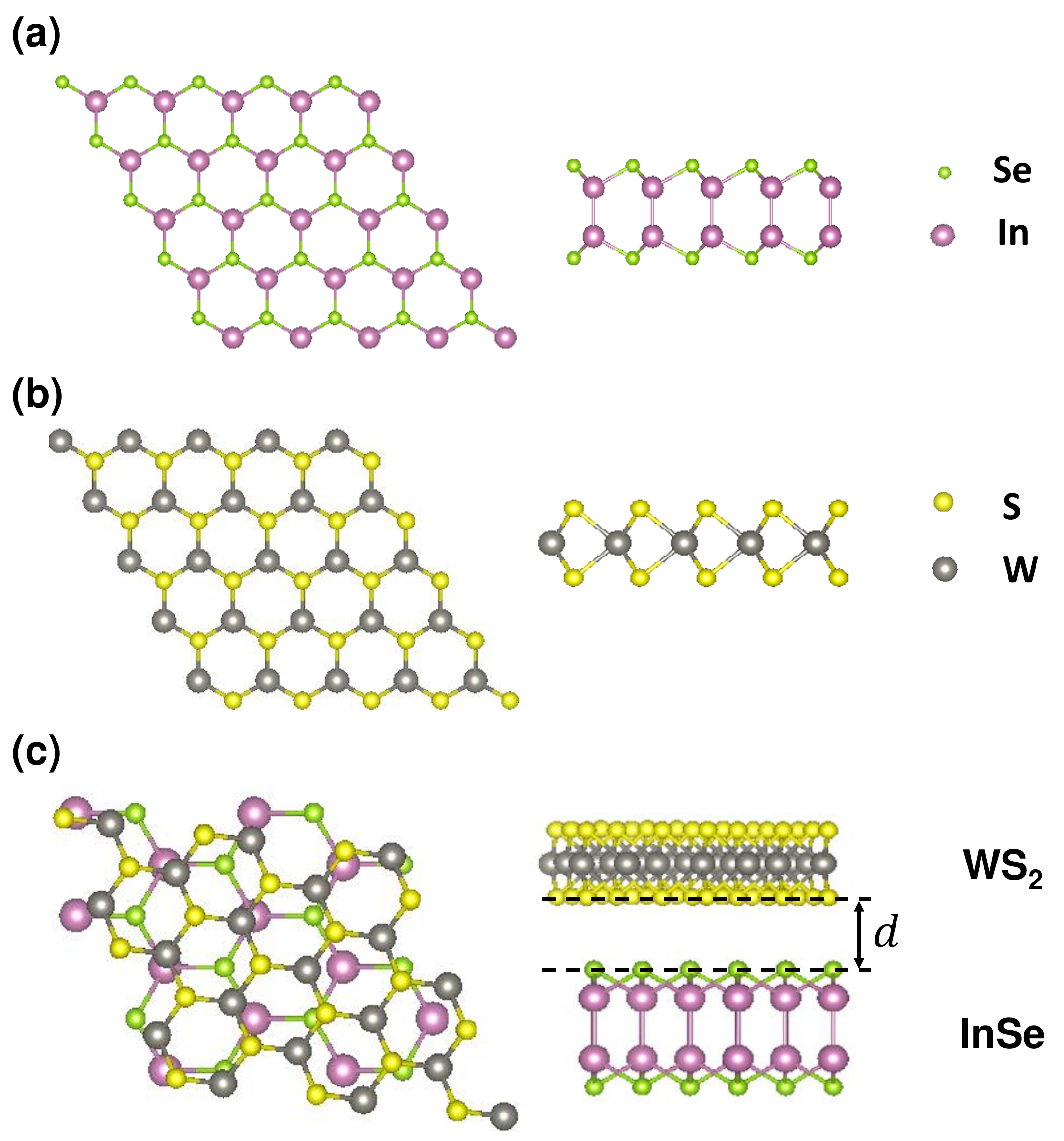}
  \caption{The top and side views of the crystal structures of (a) monolayer InSe, (b) monolayer WS$_2$, and (c) 2D InSe/WS$_2$ heterostructure. The optimized equilibrium interlayer distance is $d=3.40$ {\AA} in the heterostructure.}
  \label{fig1}
\end{figure}

The crystal structures of 2D InSe, WS$_2$, and InSe/WS$_2$ heterostructure are chosen as follows. Monolayer InSe has a honeycomb lattice connected by Se-In-In-Se sequence as shown in Fig.~\ref{fig1} (a). Monolayer WS$_2$ also has a hexagonal configuration, with each W atom anchored by three pairs of S atoms by S-W-S sequence as shown in Fig.\ref{fig1} (b). We optimized the structures of InSe and WS$_2$ monolayers to lay a foundation for the investigation on the InSe/WS$_2$ heterostructure. Our structural optimizations yield lattice constants $4.08$ {\AA} for monolayer InSe and $3.19$ {\AA} for monolayer WS$_2$. The results are listed in Table~\ref{tab1} and they are in good agreement with the values in previous reports~\cite{Zhuang2013,Kumar2018,Zhuang2013WS2,Kang2013,Amin2014,Zeng2015,Debbichi2015}. We build the 2D InSe/WS$_2$ heterostructure in such a way that the lattice mismatch is made as small as possible and the computational cost is still acceptable at the same time. Thus, a $\sqrt{12}\times\!\!\sqrt{12}$ supercell of InSe (24 indium and 24 selenium atoms) and a $\sqrt{19}\times\!\!\sqrt{19}$ supercell of WS$_2$ (19 tungsten and 38 sulfur atoms) were selected in this study as shown in Fig.~\ref{fig3} (c). The induced strains in both InSe and WS$_2$ monolayers are less than 1.7\%. Two k-point meshes, $5\times5\times1$ and $7\times7\times1$, were employed for the geometry optimizations and self-consistent calculations of the supercell of the heterostructure, respectively.

The GGA-1/2 scheme was used to make corrections to the band structures and optical properties obtained by using the usual GGA~\cite{Ferreira2008,Ferreira2011}. It is well-known that the band gaps of semiconductors are underestimated by the GGA and thus it is necessary to correct the results. Many methods have been proposed for this purpose, e.g. the GGA-1/2, GW, and HSE06 schemes. The GGA-1/2 scheme can yield results in good agreement with experiments and other theoretical methods. Meanwhile, this scheme is much less expensive than other computational methods and thus it is adopted in the present work. Within the GGA-1/2 scheme, the atomic self-energy potential is expressed as the difference between the all-electron potentials of the atom and those of the half-ion
\begin{equation}
V_s\approx{V}(0,r)-V(-1/2,r).\label{asep}
\end{equation}
The potential $V_s$ has a long-range Coulomb tail that has to be trimmed by using
\begin{equation}
\Theta(r)=
  \begin{cases}
    \left[1-\left(\dfrac{r}{CUT}\right)^m\right]^3 &,\quad r\leq{CUT}; \\
    \hfil 0 \hfil &,\quad r>CUT.
  \end{cases}
\end{equation}
The values of $m$ and $CUT$ are chosen to ensure that the result of the band gap reaches its extreme. Table IV of Ref.~\citenum{Ferreira2011} shows that, for many important semiconductors, such as Ge, Si, GaN, GaAs, and ZnO, the band gaps (0.70, 1.137, 3.52, 1.41, and 3.29~eV) resulting from the GGA-1/2 scheme are in good agreement with the GW calculations (0.66-0.83, 0.95-1.10, 3.15-3.47, 1.40-1.70, and 2.51-3.07 eV) and experimental values (0.66-0.74, 1.17, 3.507, 1.519, and 3.4 eV).

\begin{table}
\small
  \caption{The equilibrium lattice constants $a$ (in unit of {\AA}) and band gaps $E_\mathrm{g}$ (in unit of eV) of monolayer InSe and WS$_2$. As a comparison, we also list some results for the band gaps obtained by GW method, $E_\mathrm{g}$(GW), in previous works.}
  \label{tab1}
  \begin{tabular*}{0.48\textwidth}{@{\extracolsep{\fill}}lllll}
    \hline
    Materials & $a$ & $E_\mathrm{g}$(PBE) & $E_\mathrm{g}$(GGA-1/2) & $E_\mathrm{g}$(GW) \\
    \hline
    InSe & 4.08 & 1.47 & 2.75 & 2.83$^b$\\
    WS$_2$ & 3.19 & 1.77 & 2.59 & 2.64$^c$\\
    \hline
  \end{tabular*}
  $^b$From Ref.~\citenum{Zhuang2013}; $^c$From Ref.~\citenum{Zhuang2013WS2}
\end{table}

We determined the band edge positions using the equation proposed by Toroker et al.~\cite{Toroker2011}
\begin{equation}
E_\mathrm{CBM/VBM}=E_\mathrm{BGC}\pm\frac{1}{2}E_g^\mathrm{QP}
\end{equation}
where $E_\mathrm{BGC}$ denotes the band gap center energy calculated with the PBE functional and it is insensitive to different exchange functionals. Here $E_\mathrm{g}^\mathrm{QP}$ represents the quasiparticle band gap from GGA-1/2 method.

\begin{figure}[h]
\centering
  \includegraphics[width=0.48\textwidth]{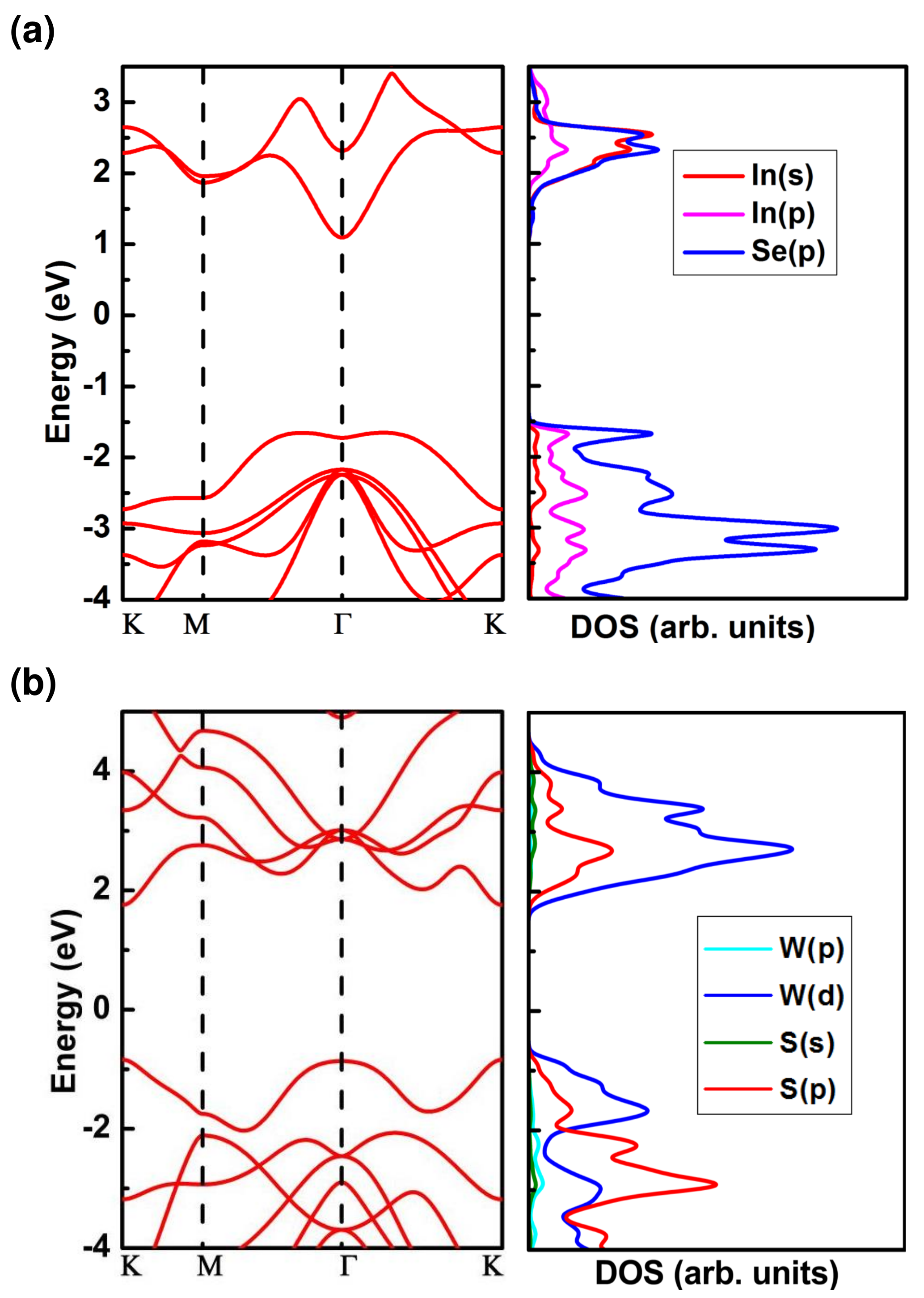}
  \caption{The band structures and DOS for monolayer (a) InSe and (b) WS$_2$.}
  \label{fig2}
\end{figure}

\begin{figure}[h]
\centering
  \includegraphics[width=0.48\textwidth]{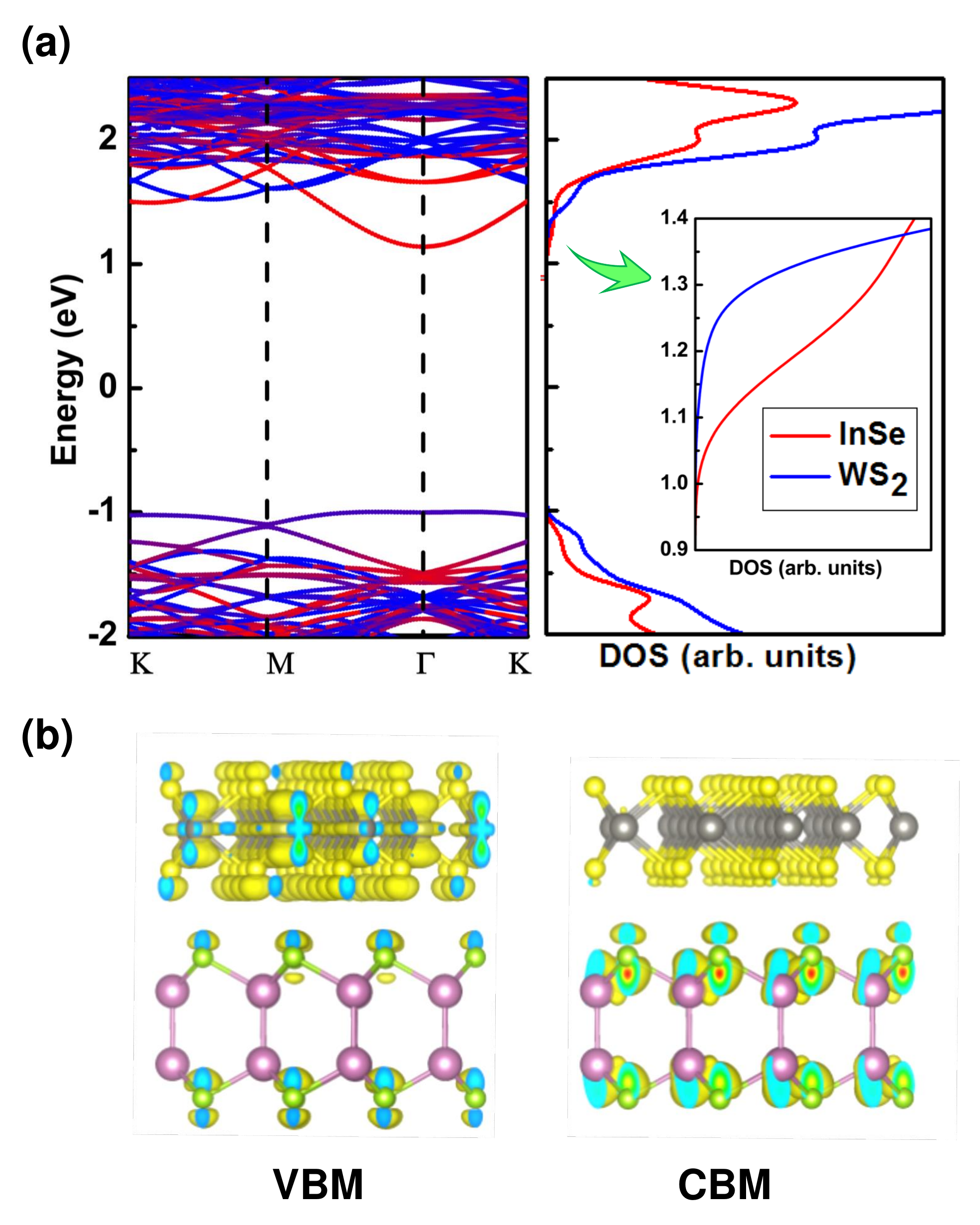}
  \caption{(a) The weighted band structure and DOS of the 2D InSe/WS$_2$ heterostructure; (b) The charge density at the VBM (CBM) depicted by the isosurface ($\rho=1.36\times10^{-4}$ $e${\AA}$^{-3}$) in the heterostructure.}
  \label{fig3}
\end{figure}

The GGA-1/2 scheme was applied first to monolayer InSe and WS$_2$ to calibrate computational parameters. In our present work, we use the following valence electron configurations: In~($4\mathrm{d}^{10}5\mathrm{S}{^2}5\mathrm{P}^1$), Se~($4\mathrm{s}^{2}4\mathrm{p}^{4}$), W~($5\mathrm{p}^{6}5\mathrm{d}^{4}6\mathrm{s}^{2}$), and S~($3\mathrm{s}^{2}3\mathrm{p}^{4}$). The half ionization is applied to the p-orbitals of the S and Se atoms. Meanwhile, we use the power $m=80$ (100) and trimming parameter $CUT=3$ (3.3) for Se (S) atoms. The band gaps calculated with these parameters are in good agreement with the GW results listed in Table~\ref{tab1}. Furthermore, the band structures and corresponding density of states (DOS) of the two monolayers are shown in Fig.~\ref{fig2}. Monolayer InSe has an indirect band gap (2.75 eV) with the CBM located at $\Gamma$ point and the VBM located between $\Gamma$ and K points as shown in Fig.~\ref{fig2}~(a). Its CBM (VBM) is mainly composed of Se 4p and In 5s (5p) orbitals, which is consistent with earlier results~\cite{Zhuang2013}. On the other hand, Fig.~\ref{fig2}~(b) shows that monolayer WS$_2$ has a direct band gap of 2.59~eV with both its CBM and VBM located at K points. Its projected DOS indicates that W 5d orbitals play the dominant role in both its CBM and VBM, while the S 3p orbitals just make a minor contribution. The electronic properties of monolayer WS$_2$ are also consistent with the results of previous calculations~\cite{Zhuang2013WS2}.

We calculated the optical absorption spectrum of the 2D InSe/WS$_2$ heterostructure as well as those of monolayer InSe and WS$_2$. The frequency-dependent dielectric function, $\epsilon(\omega)=\epsilon_1(\omega)+i\epsilon_2(\omega)$, has been calculated. The real part $\epsilon_1(\omega)$ is evaluated by the Kramers-Kronig transformation and the imaginary part $\epsilon_2(\omega)$ is calculated by summing over a large (enough) number of empty states. The absorption coefficient $I(\omega)$ is given by the following expression~\cite{Saha2000}
\begin{equation}
I(\omega)=\sqrt{2}\omega\left[\sqrt{\epsilon_1^2(\omega)+\epsilon_2^2(\omega)}-\epsilon_1(\omega)\right]^{1/2}.
\end{equation}
Taking into account of the tensor properties of the dielectric functions, we averaged $\epsilon_1(\omega)$ and $\epsilon_2(\omega)$ over three polarization vectors (along $x$, $y$, and $z$ directions).

The electron and hole mobilities were computed by the widely used formula~\cite{Ding2017,Qiao2014}
\begin{equation}\label{mobility}
\mu_\mathrm{2D}=\frac{e\hbar^3C_\mathrm{2D}}{k_\mathrm{B}Tm^\ast{m}_\mathrm{d}E_1^2},
\end{equation}
where $T$ stands for the temperature. Various effective masses are used in Eq.~(\ref{mobility}): $m^\ast$ is the one in the transport direction, i.e., $m_x^\ast$ or $m_y^\ast$, and $m_\mathrm{d}=\sqrt{m_x^\ast{m}_y^\ast}$ gives the average one over the $x$ and $y$ directions. The deformation potential constant $E_1$ is given by $E_1=\Delta{V}/(\Delta{l}/l_0)$, where $\Delta{V}$ is the change of CBM (VBM) in energy in response to proper lattice compression or dilatation described by $\Delta{l}/l_0$. Here $l_0$ stands for the lattice constant in the transport direction and $\Delta{l}$ is the deformation relative to $l_0$. The in-plane stiffness $C_\mathrm{2D}$ is defined as $C_\mathrm{2D}=2[\partial^2E/\partial(\Delta{l}/l_0)^2]/S_0$, where $E$ is the total energy of the supercell and $S_0$ denotes the area of the optimized supercell.

\section{Results and discussions on the 2D InSe/WS$_2$ heterostructure}
\label{sec3}

This section presents and discusses the results for the fundamental electronic properties, optical properties, carrier mobilities, and band edge alignment of the 2D InSe/WS$_2$ heterostructure.

The stability of the InSe/WS$_2$ heterostructure can be described by the binding energy, $E_\mathrm{b}=(E_\mathrm{InSe/WS_2}-E_\mathrm{InSe}-E_\mathrm{WS_2})/n$, where $n$ denotes the number of atoms. Moreover, $E_\mathrm{InSe/WS_2}$, $E_\mathrm{InSe}$, and $E_\mathrm{WS_2}$ are the total energy of the relaxed InSe/WS$_2$ heterostructure, monolayer InSe, and monolayer WS$_2$, respectively~\cite{Kumar2018}. The calculated binding energy is $-30$ meV/atom for the heterostructure. This indicates a moderate interaction between the 2D InSe and WS$_2$ layers, and proves that the heterostructure is stable.

Figure~\ref{fig3} (a) shows the weighted band structure and the corresponding DOS of the heterostructure. Both the CBM and the VBM are located at the $\Gamma$ point, and thus the band gap has become a direct one in the heterostructure. Moreover, the band gap of the InSe/WS$_2$ heterostructure is reduced to 2.19 eV, which is much smaller than those of monolayer InSe and WS$_2$ (see Table~\ref{tab1}). This reduction will be interpreted qualitatively by the band edge alignment below (see Fig.~\ref{fig5}).

Figure~\ref{fig3} (b) depicts the isosurface charge density at the VBM and the CBM of the heterostructure. The size of the isosurface indicates that an electron at the CBM resides almost entirely in the InSe layer. This is consistent with the band structure of the CBM and its DOS shown in Fig.~\ref{fig3} (a), where the InSe layer is definitely dominant over the WS$_2$ layer. On the other hand, a hole at the VBM exhibits more complicated behavior. It has noticeable probability to appear in the InSe layer although it resides mainly in the WS$_2$ layer. This is also in agreement with the band structure of the VBM and its DOS in Fig.~\ref{fig3} (a), where the WS$_2$ layer plays a major role and the contribution of the InSe layer is considerable. Therefore, photogenerated electron-hole pairs can be effectively separated: electrons and holes reside preferably in the InSe and WS$_2$ layers, respectively. This spatial separation in turn tends to reduce the recombination rate of the electron-hole pairs and increase their diffusion length.

\begin{table*}
\small
  \caption{The electron and hole mobilities are listed together with the associated coefficients for monolayer InSe, monolayer WS$_2$, and the 2D InSe/WS$_2$ heterostructure. Both $\mathrm{x}$ and $\mathrm{y}$ components are given for the effective masses $m^\ast$, deformation potential constants $E_1$, in-plane stiffness $C_\mathrm{2D}$, and mobilities $\mu_\mathrm{2D}$. Refer to Eq.~(\ref{mobility}) and the text following it for the detailed explanation of the various coefficients.}
  \label{tab2}
  \begin{tabular*}{\textwidth}{@{\extracolsep{\fill}}llllllllll}
    \hline
    Carriers & Materials & $m_\mathrm{x}^\ast$ & $m_\mathrm{y}^\ast$ & $E_\mathrm{1x}$ & $E_\mathrm{1y}$ & $C_\mathrm{2D-x}$ & $C_\mathrm{2D-y}$ & $\mu_\mathrm{2D-x}$ &  $\mu_\mathrm{2D-y}$ \\
    \hline
    Electrons & InSe & 0.18 & 0.19 & $-$4.42 & $-$4.32 & 49.47 & 48.70 & 1667 & 1584 \\
      & WS$_2$ & 0.45 & 0.33 & $-$11.20 & $-$11.80 & 145.79 & 148.51 & 122 & 179\\
      & InSe/WS$_2$ & 0.23 & 0.24 & $-$4.00 & $-$4.10 & 182.53 & 185.21 & 4503 & 4168\\
    Holes & InSe & 1.84 & 2.01 & $-$1.40 & $-$1.37 & 49.47 & 48.70 & 152 & 143 \\
      & WS$_2$ & 0.53 & 0.43 & $-$5.19 & $-$5.40 & 145.79 & 148.51 & 456 & 529\\
      & InSe/WS$_2$ & 1.22 & 1.15 & $-$3.48 & $-$4.23 & 182.53 & 185.21 & 223 & 162\\
    \hline
  \end{tabular*}
\end{table*}

\begin{figure}
\centering
  \includegraphics[width=0.48\textwidth]{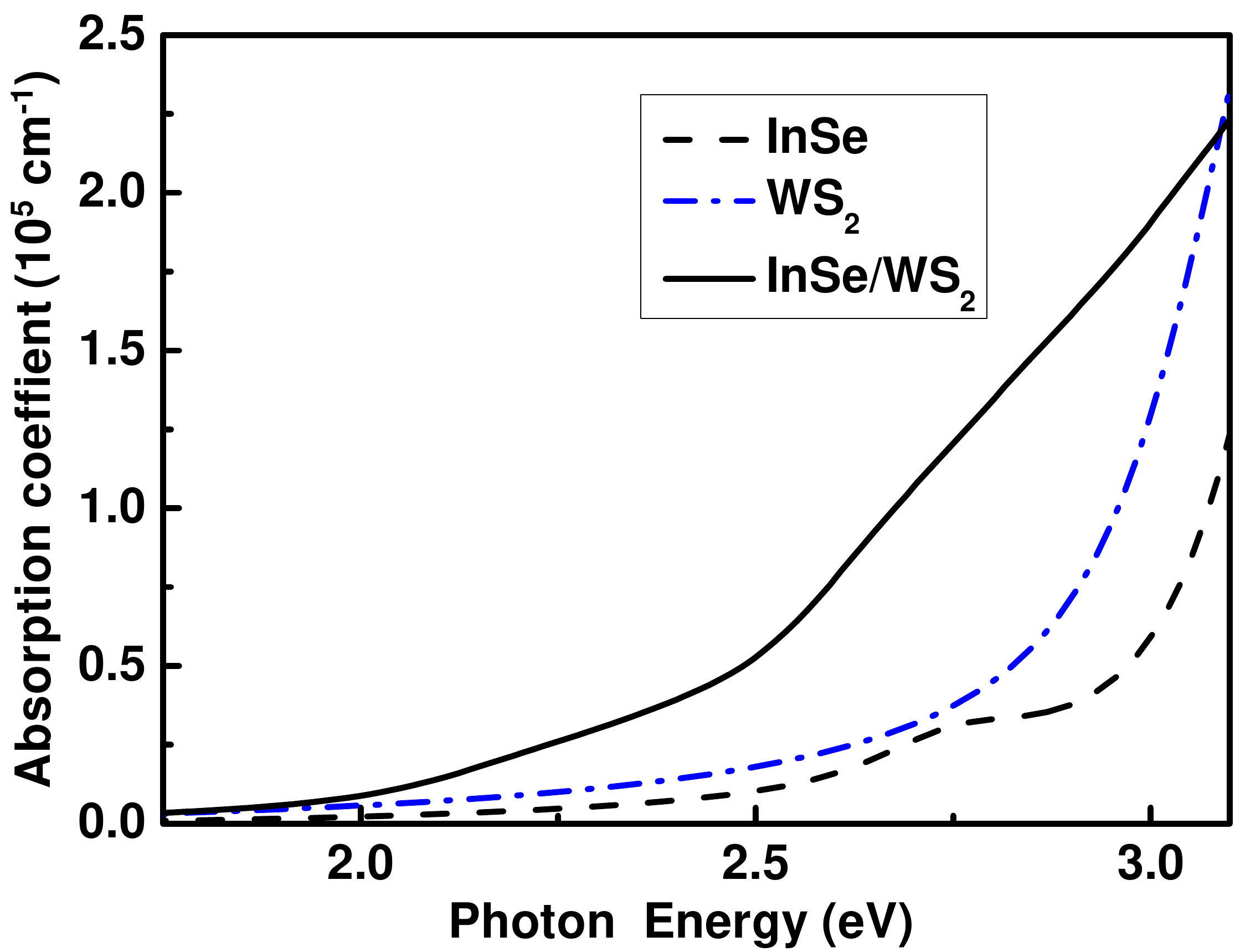}
  \caption{\textbf{Absorption spectra.} The solid-black, dashed-dotted-blue, and dashed-black curves depict the variation of the absorption coefficients with photon energy for the 2D InSe/WS$_2$ heterostructure, monolayer WS$_2$, and monolayer InSe, respectively. Only the absorption spectra for visible light are shown here.}
  \label{fig4}
\end{figure}

Figure~\ref{fig4} presents the optical absorption spectrum of the InSe/WS$_2$ heterostructure together with those of monolayer InSe and WS$_2$. The band gaps determined by fitting the absorption spectra are roughly in agreement with those given by the band structures in Figs.~\ref{fig2} and~\ref{fig3} (a). In comparison to monolayer InSe and WS$_2$, the spectrum of the heterostructure has a much wider energy region, in which the absorption coefficient for visible light is on the order of 10$^5$ cm$^{-1}$. Specifically, its absorption coefficient has a nearly fivefold (threefold) increase at the light wavelength of 490 nm in comparison to monolayer InSe (WS$_2$). Note that the wavelength 490 nm corresponds approximately to the photon energy 2.53 eV, at which the AM1.5 solar radiation spectrum reaches its maximum. The enhanced light absorption is owing to the smaller direct band gap of the heterostructure as shown in Fig.~\ref{fig3} (a).

Table~\ref{tab2} lists the electron and hole mobilities of the InSe/WS$_2$ heterostructure together with those of monolayer InSe and WS$_2$. The most significant change is the pronounced increase in the electron mobility of the heterostructure, i.e., nearly threefold increase in comparison to that of monolayer InSe. At the same time, the electron mobility of the heterostructure is much larger than that of the monolayer WS$_2$, which indicates that there is no close relationship between the two mobilities. These features can be interpreted as follows. The electron mobility is dominated by the contribution from the InSe layer since the photogenerated electrons almost reside in this layer as shown by the band structure and the CBM electron density in Fig.~\ref{fig3}. This explains why the electron mobility of the WS$_2$ is irrelevant to that of the heterostructure. Moreover, the pronounced increase in the electron mobility of the heterostructure results from the increase of the stiffness ($C_\mathrm{2D-x(y)}$) relative to that of monolayer InSe. The stiffness of the heterostructure is roughly equal to the sum of those of InSe and WS$_2$ monolayers.

As far as the hole mobility is concerned, the situation is more complicated than that of the electrons. The hole mobility of the heterostructure increases remarkably in comparison to that of monolayer InSe while it decreases significantly as compared with that of  monolayer WS$_2$. These mobilities are of the same order although they have different values. This suggests that both the InSe and the WS$_2$ layers have considerable contributions to the hole mobility of the heterostructure. This can be interpreted as follows. The photogenerated holes reside mostly in the WS$_2$ layer and partly in the InSe layer as shown by the band structure and VBM electron density in Fig.~\ref{fig3}. This is consistent with the change in the hole effective mass of the heterostructure shown in Table~\ref{tab2}, i.e., the hole effective mass falls between the values of the InSe and the WS$_2$ layers. Meanwhile, $E_\mathrm{1x(y)}$ also exhibits the similar variation. As compared with monolayer InSe, $m_\mathrm{x(y)}^\ast$ of the heterostructure decreases and tends to increase the hole mobility, while $E_\mathrm{1x(y)}$ increases in magnitude and makes the hole mobility smaller instead. The two variations cancel each other partly in the contribution to the hole mobility. However, the hole mobility has an overall increase in comparison to that of the InSe layer when the large increase in the stiffness is taken into account.

\begin{figure}[h]
\centering
  \includegraphics[width=0.48\textwidth]{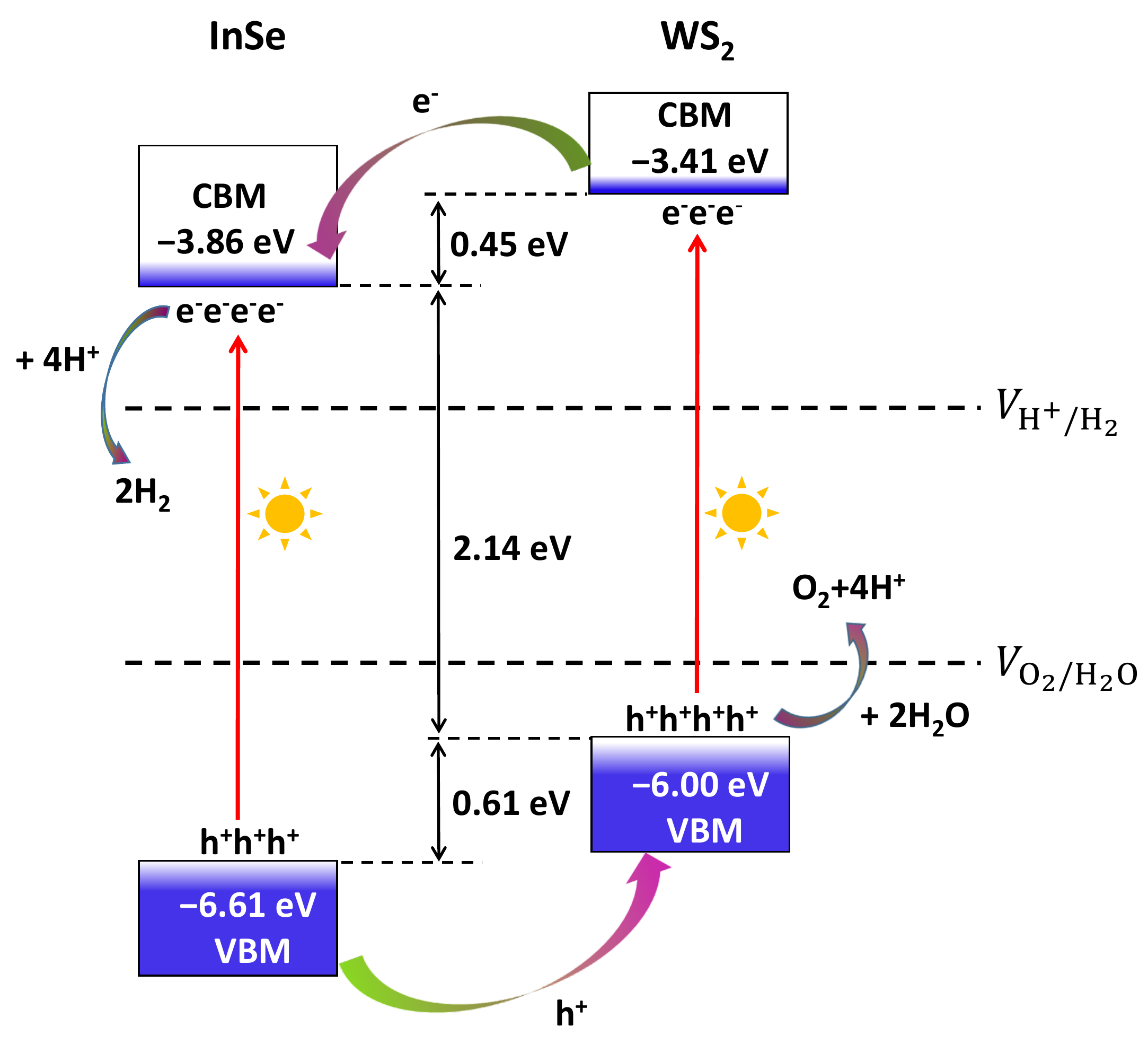}
  \caption{The band edge positions of monolayer InSe and WS$_2$ are plotted together with the oxidation ($V_{\mathrm{O}_2/\mathrm{H}_2\mathrm{O}}=-5.67$ eV) and reduction ($V_{\mathrm{H}^+/\mathrm{H}_2}=-4.44$ eV) potentials of water relative to the vacuum level. The photocatalytic water splitting is also illustrated schematically using the InSe/WS$_2$ heterostructure.}
  \label{fig5}
\end{figure}

Figure~\ref{fig5} shows the band edge positions of monolayer InSe and WS$_2$ relative to the vacuum level. The CBM and VBM of monolayer InSe are $-3.86$ eV and $-6.61$ eV, respectively, which are in agreement with previous results~\cite{Zhuang2013}. As for monolayer WS$_2$, the CBM and VBM are $-3.41$ eV and $-6.00$ eV, respectively, which are roughly consistent with previous reports~\cite{Zhuang2013WS2}. Both the CBM and the VBM of InSe are lower than those of WS$_2$. The conduction band offset (CBO) and valence band offset (VBO) are 0.45 eV and 0.61 eV, respectively. Thus the InSe/WS$_2$ heterostructure has a typical type-II band alignment, which is consistent with the electronic structures in Fig.~\ref{fig3}. The overall band gap of the heterostructure is reduced to 2.14 eV as shown by the up-down arrow between the VBM of WS$_2$ and the CBM of InSe. This band gap is close to the more rigorous result (2.19 eV) given by the band structure in Fig.~\ref{fig3} (a). The difference is attributed to the strain and interlayer interaction in the heterostructure.

Figure~\ref{fig5} also illustrates schematically the photocatalytic water splitting based on the heterostructure. A basic criterion whether a photocatalyst can be used for water splitting is that the redox potentials of water must lie between its CBM and VBM. It is well-established that $V_{\mathrm{H}^+/\mathrm{H}_2}$ and $V_{\mathrm{O}_2/\mathrm{H}_2\mathrm{O}}$ are $-4.44$~eV and $-5.67$~eV, respectively, relative to the vacuum level~\cite{Zhuang2013}. Thus the CBM of monolayer InSe is $0.58$ eV higher than $V_{\mathrm{H}^+/\mathrm{H}_2}$, and its VBM is $0.94$ eV lower than $V_{\mathrm{O}_2/\mathrm{H}_2\mathrm{O}}$. Meanwhile, the CBM of monolayer WS$_2$ is $1.03$ eV higher than $V_{\mathrm{H}^+/\mathrm{H}_2}$, and its VBM is $0.33$ eV lower than $V_{\mathrm{O}_2/\mathrm{H}_2\mathrm{O}}$. Thus both InSe and WS$_2$ monolayers satisfy the basic requirement for water splitting. On the other hand, the CBM of the InSe/WS$_2$ heterostructure is $0.58$ eV higher than $V_{\mathrm{H}^+/\mathrm{H}_2}$, and its VBM is $0.33$~eV lower than $V_{\mathrm{O}_2/\mathrm{H}_2\mathrm{O}}$. This indicates that the heterostructure still satisfies the requirement although its band gap has been reduced. The existence of the CBO drives the photogenerated electrons in the CBM of the WS$_2$ layer to the CBM of the InSe layer. Then these electrons can take part in the reduction reaction to produce H$_2$ as illustrated by Fig.~\ref{fig5}. Similarly, the existence of VBO drives the photogenerated holes in the VBM of the InSe layer to the VBM of the WS$_2$ layer, and then these holes participate in the oxidation reaction producing O$_2$. Moreover, the separation of the electron-hole pairs can also effectively restrains their recombination and in turn increases their diffusion length. The enhanced carrier mobilities of the heterostructure also help the electrons and holes move faster to the active sites and thus improve the catalytic performance.

\section{Conclusions}
\label{sec4}

In summary, we designed a 2D vdW heterostructure composed of two monolayers, InSe and WS$_2$, with the purpose to enhance its optoelectronic performance in the visible region. First-principles calculations have been employed to investigate the fundamental electronic properties, optical properties, carrier mobilities, and basic photocatalytic characteristics of the heterostructure. Our results show the following desirable properties of the heterostructure. First, it has a type-II band alignment, i.e., both the CBM and the VBM of InSe are lower than those of WS$_2$. Consequently, its band gap is reduced to 2.19 eV, which is much smaller than those of monolayer InSe and WS$_2$. Its CBM and VBM derive mainly from the InSe and WS$_2$ layers, respectively, which promotes effectively the spatial separation of the photogenerated electron-hole pairs and in turn decreases their recombination rate. Second, the visible-light absorption is increased considerably, e.g. about fivefold (threefold) increase at the wavelength of $490$ nm in comparison to monolayer InSe (WS$_2$). Third, the two monolayers complement each other in carrier mobilities: the photo-generated electrons and holes inherit the large mobilities from InSe and WS$_2$ monolayers, respectively. Last but not least, the VBM and CBM of the heterostructure still straddle properly the oxidation and reduction potentials of water, which satisfies the basic requirement of photocatalytic water splitting. These results suggest strongly that the InSe/WS$_2$ heterostructure and multilayered structures based on it are highly promising for optoelectronic and photoelectrochemical applications.

\section*{Acknowledgements}
This work was supported by the National Natural Science Foundation of China (11404013, 61705003, 11605003, 11474012, 11364030, 61622406, 61571415, and 51502283) and the National Key Research and Development Program of China (Grant No. 2017YFA0206303, MOST of China). We used computational resource of the ``Explorer 100" cluster system of Tsinghua National Laboratory for Information Science and Technology and part of the analysis was performed on the Computing Platform of the Center for Life Science of Peking University.



\bibliographystyle{elsarticle-num}
\bibliography{InSe-WS2}

\end{document}